# A Meta-Learning Approach for Software Refactoring

Hanieh Khosravi[a], Abbas Rasoolzadegan[a,1]

[a] Faculty of Engineering, Ferdowsi University of Mashhad, Mashhad, Iran

## Abstract

Software refactoring is the process of changing the structure of software without any alteration in its behavior and functionality. Presuming it is carried out in appropriate opportunities, refactoring enhances software quality characteristics such as maintainability and extensibility. Thus far, various studies have addressed the problem of detecting proper opportunities for refactoring. Most of them are based on human expertise and are prone to error and non-meticulous. Fortunately, in recent efforts, machine learning methods have produced outstanding results in finding appropriate opportunities for refactoring. Sad to say, Machine learning methods mostly need plenty of data and, consequently, long processing time. Furthermore, there needs to be more annotated data for many types of refactoring, and data collection is time-consuming and costly. Accordingly, in this paper, we have formulated the problem of detecting appropriate opportunities for refactoring as a few-shot classification problem. We have utilized model-agnostic meta-learning (MAML), a recognized meta-learning algorithm, to learn a neural network on tasks from high-resource data. The trained model, then, is adapted to a model with high accuracy for tasks from low-resource data. Experimental results revealed 91% accuracy, which illustrates the effectiveness and competitiveness of our proposed meta-learning model.

## 1. Introduction

It is inevitable that requirements change during the software life cycle [1]. As a result, the design and internal structure of the software should be modified to adapt to new requirements. Desirable quality characteristics of software design results in less time and cost consumption for adaptation. Poor software design detracts from the final product's performance and adversely affects the development process [2]. In such circumstances, software refactoring is presented as an excellent idea for the amendment of software design.

Software refactoring is the process of changing the structure and design of software, aiming to enhance its qualitative characteristics without altering its behavior [3, 4]. During the software life cycle, the expansion of requirements and scaling up of software encourage developers to perform the refactoring process [5].

Therefore, refactoring can be considered a part of the software development process [6]. Considering the fact that the quality of the software is assessed by the degree of satisfaction of stakeholders' needs [7], refactoring improves the quality of software by expediting and facilitating the process of responding to requirements. Refactoring enhances qualitative characteristics such as understandability, maintainability, reusability, and flexibility of software [8], provided that it is carried out in proper opportunities.

So far, numerous methods have been proposed for the problem of detecting appropriate opportunities for refactoring. A significant part is based on heuristics, human expertise, and manual checking of code (or design) properties [9-13]. Such methods are error-prone, and some relations and decisive symptoms may be mistakenly ignored. However, learning-based methods have recently surpassed existing ones and used a mathematical point of view to produce highly accurate results [14-26]. One is the research carried out by Aniche et al. [26]. They trained six machine learning algorithms with a dataset consisting of about two million data for twenty different types of refactoring and achieved an excellent accuracy. Even so, their model has two restrictions: 1) requiring a large amount of data for each refactoring (there is not a large annotated dataset for many types of refactoring), and 2) training a distinct model for each refactoring from scratch.

To remedy the first deficiency, we framed the problem of discovering appropriate opportunities for refactoring as a few-shot learning problem. Few-shot learning is the problem of training a model with very few annotated data. To alleviate the second limitation, we exploited meta-learning to train our model. Meta-learning gives our model the noteworthy property of fast adaptation of a meta-trained model to a precise model for a new task (with few data) without requiring training from scratch. Meta-learning proposes a prominent solution for the problem of few-shot learning. In a nutshell, Meta-learning tries to model the manner of learning procedure of human beings, i.e., quickly learning new concepts based on previous knowledge and a small number of input observations about the new concepts [27].

In this paper, we used MAML [27], a famous meta-learning algorithm. The learning procedure composes of two nested levels. At the outer level, a base learner accumulates knowledge from a crowd of similar tasks. The accumulated knowledge actually acts as a prior knowledge (or meta-knowledge) for the inner level of learning. At the inner level, a meta-learner tries to quickly learn generic concepts from a new task with the aid of and based on prior knowledge. After accomplishing the learning procedure, known as the meta-training stage, the meta-testing stage will be executed. In this stage, the trained model adapts to a model with high accuracy for an unseen task with very few data.

In recent years, meta-learning has attained considerable success in various domains and applications, including medical image processing [28], medical diagnostics [29], learning optimizers[30], aerial scene classification [31], neural language processing [32], semi-supervised few-shot classification [33], urban traffic prediction [34], recommendation systems [35], drug discovery [36], continual learning [37], Noise identification [38], object detection (instance detection) [39], computer vision [40], online learning [41], Reinforcement Learning [42], semantic segmentation [43]. Because meta-learning obviates the need for a large amount of annotated data, the costly and tedious process of collecting annotated data, long processing time, powerful processors, and high memory capacity.

The remainder of the paper is arranged as follows: a review of conducted research in discovering appropriate opportunities for refactoring is presented in Section 2. Section 3 looks at some basic concepts and preliminary about few-shot classification and meta-learning. Our proposed meta-learning model is clearly explained in Section 4, and Section 5 presents experimental validation and a discussion of results. Finally, in Section 6, we discuss conclusions and future works.

## 2. Related Work

Yue et al. [14] have introduced a learning-based approach for recommending clones based on extracted features concerning software projects' present and past status. They extracted 34 features and trained an AdaBoost model. Their conducted evaluation on six projects achieved 76% and 83% of F-scores in cross-project and within-project settings for clone refactoring recommendations. Clone detection has been investigated in another research by Wang et al. [15]. In their study, a decision tree-based classifier was trained using 15 clone-related metrics. Nyamawe et al. [16] have also explored previously applied 14 types of refactoring in 43 java open-source projects and then trained two classifiers: the first determines whether refactoring is required or not, and the second is a multi-label classifier that suggests all needed refactoring types. They have trained Linear Regression, Random Forest, Multinomial Naive Bayes, and Support Vector Machine (SVM) classifiers that revealed 73% of average precision.

Nyamawe et al. [17] approach suggests refactorings using the history of previously applied refactorings, requested features, and code smells. They have used a variety of classifiers such as Logistic Regression, Convolutional Neural Network, Multinomial Naïve Bayes, Decision Tree, Random Forest, and SVM. For predicting commits with refactoring operations, Sagar et al. [18] explored 800 open-source java projects to pick out 5004 commits from all extracted commits and create a balanced dataset of five types of refactoring. They have used Random Forests, SVM, and Linear Regressions to identify appropriate refactoring opportunities.

Kumar et al. [19] have utilized Least Squares Support Vector Machines (LS-SVM) to detect class-level refactoring candidates. Initially, they selected 102 source code metrics and then utilized PCA to degrade them to 31 features. LS-SVM RBF kernels have achieved an average AUC of 0.96 in their evaluations. In another research, Kumar et al. [20] computed 25 method-level source code quality metrics, such as McCabe's cyclomatic complexity, clone classes, and total lines of code, to identify methods that require refactoring using ten different classifiers. Their experiments have revealed 98.16% and 98.17% of mean accuracy for AdaBoost and ANN+GD classifiers, respectively. Kosker et al. [21] also calculated 26 code complexity metrics containing Halstead metrics, McCabe's cyclomatic complexity, and lines of code for four versions of one project. By using Weighted Naive Bayes, they were able to identify 82% of classes that required refactoring.

Model-level refactoring has also been studied by Sidhu et al. [22]. To detect refactoring opportunities using a deep neural network, they have gathered a set of metrics for calculating design characteristics, including size, coupling, inheritance, modularity, and encapsulation. Their model has obtained a precision, recall, and F1 score, respectively, equal to 87%, 93%, and 90%.

Kurbatova et al. [23] have proposed a method to suggest move method refactoring utilizing SVM. The input data of SVM are embeddings generated for each method of code using a path-based representation of code called code2vec and contain syntactic and semantic information. SVM has also been employed by Akour et al. [24] for the detection of class-level refactoring opportunities. They have exerted SVM along with genetic and whale optimization algorithms on a dataset of four open-source projects.

Xu et al. [25] have suggested a method for the recommendation of extract method refactoring using gradient boosting. Both structural and behavioral information is extracted to encode cohesion, coupling, and complexity concepts. Aniche et al. [26] have used Neural Network, Random Forest, Decision Tree, Naïve Bayes, SVM, and Logistic Regression to detect appropriate opportunities for 20 types of refactoring. They have generated a dataset comprising code, process, and ownership features of refactorings committed on 11,149 extensive Java projects obtained from GitHub, Apache, and F-Droid sources.

## 3. Preliminary

This section presents some preliminary and basic concepts about few-shot classification and meta-learning that facilitate better comprehension of the rest of the paper.

Let $D = \{(x, y)_i\}_{i=1}^{n}$ represents a dataset, where $(x, y)_i$ specifies a pair of data and its equivalent label. The dataset contains abundant instances for some classes and scarce instances for others. We split $D$ into meta-train ($D_{meta-train}$) and meta-test ($D_{meta-test}$) sets. There is no overlap between the classes of the two sets. $D_{meta-train}$ encompasses classes with abundant data, and $D_{meta-test}$ comprises classes with scarce data. The goal is to train the model's parameters on $D_{meta-train}$ and fine-tune to $D_{meta-test}$.

In the context of few-shot learning, a model is often trained and evaluated on a series of N-way K-shot classification tasks (or episodes). Each task is constituted of two splits: support-set and query-set. The support set holds N classes and K labeled samples of each class. Therefore, the support-set contains N × K samples for learning the task. The query-set also contains the same N classes and Q samples of each one. The query-set includes N × Q data samples that evaluate the performance of the trained model on the current task. Commonly, Q is greater than K. It is noteworthy that the whole of a task in few-shot learning acts as a training data point in conventional machine learning. Figure 1 demonstrates the construction of the aforementioned datasets and tasks.

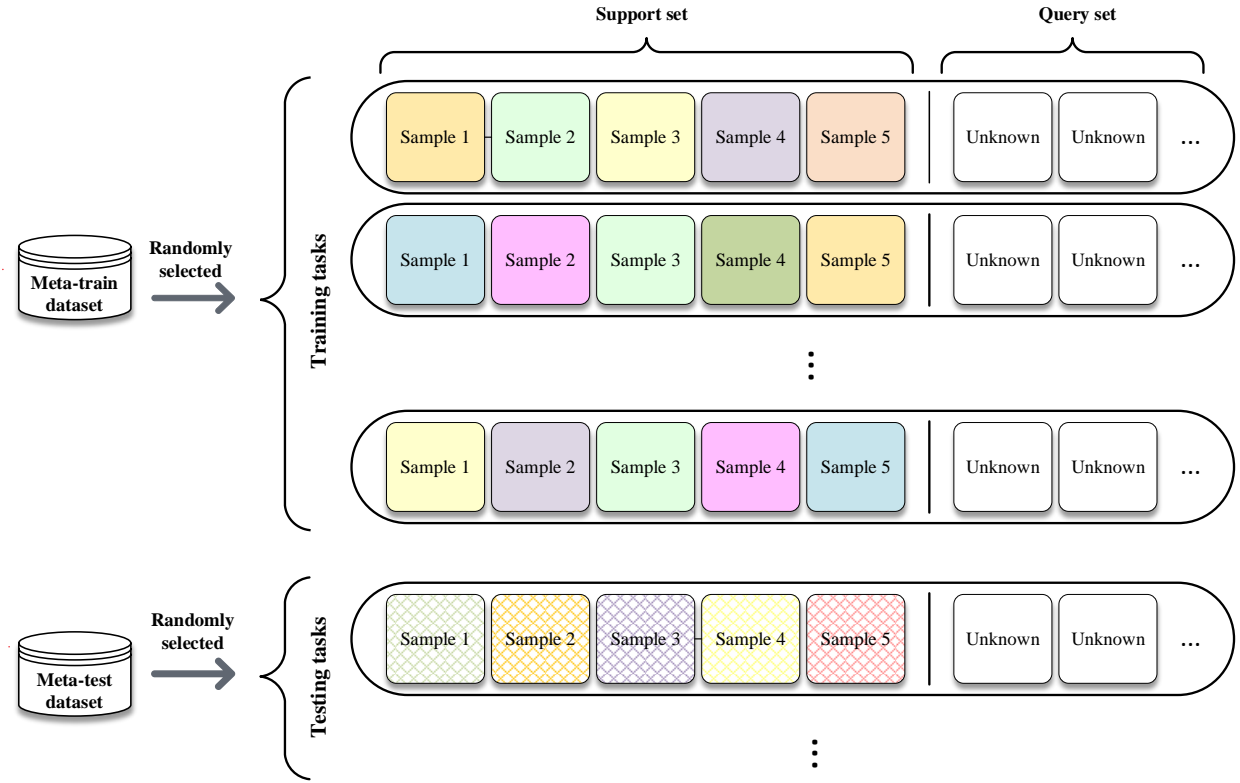

**Figure 1.** The above figure illustrates two sets of tasks: training and testing tasks (respectively drawn from the meta-train and meta-test datasets). The tasks are 5-way 1-shot and each color represents a distinct class (or label). The support-set of each task contains 5 classes and 1 sample of each one (totally 5 samples) and the query-set holds the same 5 classes and 3 samples of each class (totally 15 samples). The labels of query samples are unknown to the model and are utilized to evaluate the trained model.

## 4. Methodology

In this section, we will describe our proposed method in detail. First, we look at how to cleave the dataset into $D_{meta-train}$ and $D_{meta-test}$ sets and create training and testing tasks, and then, the MAML algorithm for few-shot meta-learning and its special features will be explained.

### 4.1. Creation of Few-Shot Tasks

In meta-learning, accumulated knowledge from a bundle of tasks is adapted to new tasks with few samples and few learning iterations. In other words, the collected knowledge from abundant sampled tasks (training tasks) is transferred to learn new tasks with scarce data (testing tasks). The transferred knowledge is, in fact, a meta-knowledge that is shared by all of the tasks. Therefore, it is presumed that meta-learning is carried out on a distribution of related tasks. Consequently, in software refactoring, it is pragmatic to train our model on tasks created from some types of refactoring with abundant data and then fine-tune it to other types with scarce data.

**Table 1.** Refactoring types that their instances are gathered in Aniche et al. dataset.

| | Refactoring | Problem | Solution |
|---|---|---|---|
| **Class-level** | Extract Class | When a single class is responsible for multiple tasks. | Move related methods and fields to each task to a new class. |
| | Extract Subclass | When a class contains a particular use case that its related methods and fields are rarely used in specific scenarios. | Move the specific scenarios into a newly created subclass. |
| | Extract Super-class | When several classes contain common members. | Create a new super-class and move common properties to it. |
| | Extract Interface | When several classes share a common interface. | Create an interface and move common parts to it. |
| | Move Class | When the degree of relevance of a class to classes of a particular package is more than its current package. | Move the class to the relevant package. |
| | Rename Class | When the name of a class is not illustrative enough of its functionality. | Change the class name to something more descriptive. |
| | Move and Rename Class | When facing two problems of move class and rename class at the same time. | Move the class to a new package and rename it. |
| **Method-level** | Extract Method | When statements of a single method can be grouped into related bundles. | Extract each group to a new method. |
| | Inline Method | When the body of a method is more obvious than the method itself. | Replace all calls to the method with its content and remove the method itself. |
| | Move Method | When the degree of relevance of a method to a specific class is more than its current class. | Move the method to the relevant class. |
| | Pull Up Method | When several classes in an inheritance hierarchy contain a method with a similar task. | Pull the method up in the inheritance hierarchy. |
| | Push Down Method | When a behavior implemented in a super-class is utilized in one or a few sub-classes. | Push the method down in the inheritance hierarchy. |
| | Rename Method | When the name of a method is not expressive enough of its aim. | Change the method name to something more descriptive. |
| | Extract And Move Method | When facing two problems of extract method and move method simultaneously. | Extract a method and move it to a more relevant class. |
| **Variable-level** | Extract Variable | When an expression is hard to understand or is duplicated several times throughout the code. | Place the result of the expression or its parts in separate variables that are self-explanatory. |
| | Inline Variable | When a dispensable variable is defined and stayed unchanged. | Replace all occurrences of the variable in the code with its initializer. |
| | Parameterize Variable | When it is preferred that a variable becomes a method parameter. | Remove the variable declaration from the method body and put it into the method parameters. |
| | Rename Parameter | When the name of a parameter is not descriptive enough of its purpose. | Change the parameter name to something more expressive. |
| | Rename Variable | When the name of a variable is not descriptive enough of its purpose. | Change the variable name to something more expressive. |
| | Replace Variable/Attribute | When a variable/attribute is used in multiple methods. | Convert the variable/attribute to class attribute. |

In this paper, we have used the dataset created and introduced by Aniche et al. In this dataset, labeled instances of 20 types of refactoring in class, method, and variable levels are collected as mentioned and described in Table 1. The dataset encompasses 2,086,898 refactored instances, and features are values of three types of metrics: source code, process, and code ownership. Source code metrics measure quantitative attributes of code, process metrics are beneficial in defect prediction algorithms, and ownership metrics estimate quantitative details about authors and developers of code.

To construct few-shot tasks, we have cleaved the dataset into two portions, $D_{meta-train}$ and $D_{meta-test}$, as explained in section 3. Five classes of refactoring with the least instances (i.e., move and rename class, rename class, extract and move method, extract subclass, and extract variable) constitute the $D_{meta-test}$, and the others (that contain ample data for training) have been placed in $D_{meta-train}$. Instances of each dataset were shuffled to create testing and training tasks, respectively. In the next step, the tasks are fed to few-shot meta-learning algorithm that is described in the next section.

### 4.2. Few-Shot Meta-Learning

Algorithm 1 describes few-shot meta-learning with MAML. The learning procedure contains two phases: meta-train and meta-test (or fine-tuning).

**Meta-train phase:** This stage consists of two nested loops. At the inner loop (lines 4-7), a base learner denoted by $y = f(x, \theta)$ is trained for each task selected from a batch of tasks. The base learners learn related concepts to each task. To evaluate the value of the loss function, we use cross-entropy loss, represented in Equation 1:

$$\mathcal{L}_{T_i}(f_\phi) = \sum_{x^{(i)}, y^{(i)} \sim T_i} y^{(i)} \log f_\phi(x^{(i)}) + (1 - y^{(i)}) \log(1 - f_\phi(x^{(i)})) \tag{1}$$

Where $x^{(i)}$ and $y^{(i)}$ are input and label pairs sampled from $the\ i^{th}$ task (i.e., $T_i$). The parameter vector $\theta_i$ of each task is first initialized by meta-parameter $\theta$ and then updated with gradient descent and $\alpha$ step size (line 6). At the outer loop (lines 2-9) meta-learner accumulates knowledge from learned tasks, i.e., meta-parameter $\theta$ is updated from learned tasks parameters $\theta_i$ using cross-entropy loss and $\beta$ step size.

---

**Algorithm 1.** Few-shot meta-learning algorithm with MAML

| //meta-train phase | //meta-test phase |
|---|---|
| **Require:** $D_{meta-train}$: meta-train dataset | **Require:** $D_{meta-test}$: meta-test dataset |
| $\quad\quad\alpha, \beta$: step size | $\quad\quad\alpha$: step size |
| 1: randomly initialize $\theta$ | 1: **for** iteration = 1, 2, … K **do** |
| 2: **for** iteration = 1, 2, … N **do** | 2: $\quad$ Sample batch of p tasks $T_j \sim D_{meta-test}$ |
| 3: $\quad$ Sample batch of m tasks $T_i \sim D_{meta-train}$ | 3: $\quad$ Initialize mode parameter vector with |
| 4: $\quad$ **for all** $T_i$ **do** | $\quad\quad$ meta-parameter $\theta$ |
| 5: $\quad\quad$ Evaluate $\mathcal{L}_{T_i}(f_\theta)$ with cross-entropy | 4: $\quad$ **for** iteration = 1, 2, … L **do** |
| $\quad\quad\quad$ function in Equation (1) | 5: $\quad\quad$ Evaluate $\mathcal{L}_{T_j}(f_\theta)$ with cross-entropy |
| 6: $\quad\quad$ Update task parameter $\theta_i$ with | $\quad\quad\quad$ function and support samples of $T_j$ |
| $\quad\quad\quad$ gradient descent: | 6: $\quad\quad$ Update meta-parameter $\theta$ with |
| $\quad\quad\quad\theta_i = \theta - \alpha\nabla_\theta\mathcal{L}_{T_i}(f_\theta)$ | $\quad\quad\quad$ gradient descent: |
| 7: $\quad$ **end for** | $\quad\quad\quad\theta' = \theta - \alpha\nabla_\theta\mathcal{L}_{T_j}(f_\theta)$ |
| 8: $\quad$ Update meta parameter $\theta$ using task | 7: $\quad$ **end for** |
| $\quad\quad$ parameters: | 8: $\quad$ Calculate accuracy on query samples of $T_j$ |
| $\quad\quad\theta = \theta - \beta\nabla_\theta\sum_{T_i}\mathcal{L}_{T_i}(f_{\theta_i})$ | 9: **end for** |
| 9: **end for** | |

**Meta-test (fine-tuning) phase:** To fine-tune the trained meta-learner to unseen tasks with a minor number of iterations, N-way K-shot tasks are randomly sampled from $D_{meta-test}$ (line 2). The meta-parameter $\theta$ of the meta-learner is adapted to parameter vector $\theta'$ using support samples of test tasks and cross-entropy loss (lines 4-7). Afterward, the adapted parameter $\theta'$ is utilized to calculate the accuracy of the fine-tuned model on query samples (line 8).

Figure 2 also depicts the aforesaid algorithm graphically. The upper box illustrates the meta-train and the bottom demonstrates the meta-test phase. Evidently, in the meta-train phase, the meta-parameter $\theta$ is randomly initialized at first, and then in each iteration, the task parameters $\theta_i$ are updated from the initial point $\theta$. In the further step, meta-parameter $\theta$ is calculated and updated with task parameters, training tasks, and cross-entropy loss. This flow is repeated for $N$ iterations. The output of the meta-train stage is meta-parameter $\theta$, which is also the meta-test stage input. In the meta-test stage (the bottom box), the fine-tuned parameter $\theta'$ is updated from the initial point $\theta$ using testing tasks and cross-entropy loss function. A model with fine-tuned parameter $\theta'$ can classify new data in the same categories as testing tasks.

In previous machine learning-based methods in refactoring, trained models are merely capable of classifying refactoring types that participate in training data, i.e., to classify other refactorings, a new model must be trained from scratch with a plentiful amount of data of such refactorings. As previously stated, for many types of refactoring, an ample amount of data is unavailable, or data collection is complicated. According to preceding technical discussions, a magnificent property of our model is the potentiality of fine-tuning to a model for classifying newcomer types of refactoring that did not participate in training data, i.e., refactorings other than is aforesaid in Table 1. Another striking property of our model is the need for a few data for fine-tuning. In a nutshell, our method can train a model on a few types of refactoring and fine-tune it to classify many other types with few data.

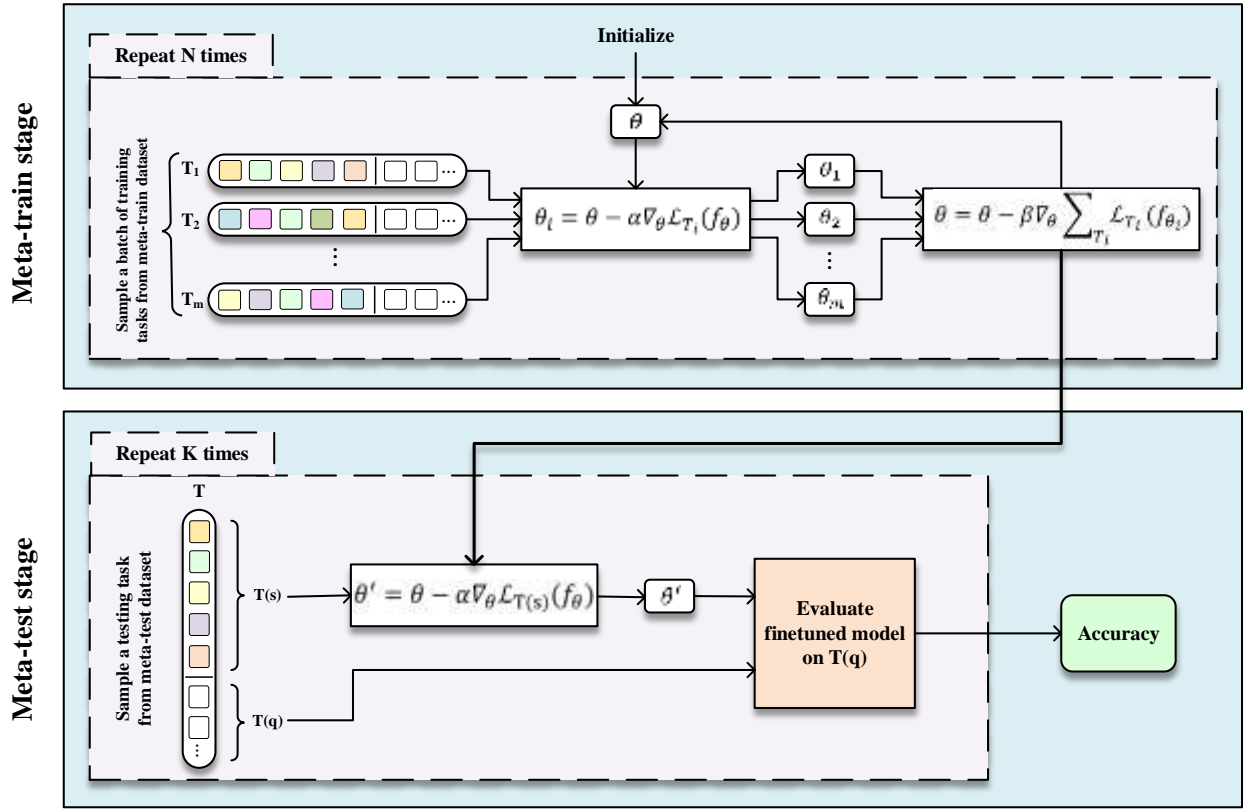

**Figure 2.** Few-shot meta-learning with MAML at a glance

## 5. Experimental Results

In this section, we will explain the experimental setting, the dataset's characteristics, and conducted experiments to evaluate and analogize our proposed model with existing ones. The results clarify the superiority of our meta-learning-based model.

### 5.1. Experiment Settings

We implemented a Multilayer Perceptron (MLP) with three hidden layers, each is instituted of 80 neurons (in our experiments, this setting has disclosed the best results). We have utilized Stochastic Gradient Descent (SGD) as optimizer for the optimization of both inner and outer optimization problems of the meta-learning algorithm, and the values of learning rates for both levels ($\alpha$ and $\beta$) are equal to 0.001. To evaluate the value of model parameters in each step of SGD, we have used the cross-entropy loss function (Equation 1) for both the meta-train and meta-test stages. The meta-train stage has been executed for 5,000 meta-iterations on batches of 25 tasks in each iteration. Meta-test stage also has run for just one adaptation step on batches of 25 tasks for each step.

### 5.2. Dataset Description

As already stated, in this paper, we have trained our model on Aniche et al. dataset. The dataset encompasses code, process, and ownership features of 20 types of refactoring committed on 11,149 very large Java projects. The Java projects are gathered from GitHub, Apache, and F-Droid sources. The number of committed refactorings detected from each source is reported in Table 2. The last column also indicates the total number of refactoring samples in all three sources. The dataset entirely comprises 2,086,898 refactoring instances extracted from all sources.

**Table 2.** Number of instances of each refactoring type identified in each project

| | Refactoring | GitHub | Apache | F-Droid | All |
|---|---|---|---|---|---|
| $D_{meta-train}$ | Extract interface | 7775 | 2363 | 357 | 10495 |
| | Extract super-class | 20027 | 5228 | 1559 | 26814 |
| | Extract class | 31729 | 6658 | 2804 | 41191 |
| | Move class | 32259 | 16413 | 1143 | 49815 |
| | Extract method | 243011 | 61280 | 23202 | 327493 |
| | Move method | 124411 | 26592 | 12075 | 163078 |
| | Inline method | 40087 | 10027 | 3713 | 53827 |
| | Push down method | 47767 | 12933 | 1930 | 62630 |
| | Pull up method | 116953 | 32646 | 5477 | 155076 |
| | Rename method | 340304 | 65667 | 21964 | 427935 |
| | Inline variable | 23126 | 5616 | 2152 | 30894 |
| | Rename Variable | 250076 | 57086 | 17793 | 324955 |
| | Rename Parameter | 261186 | 61246 | 14319 | 336751 |
| | Parameterize Variable | 16542 | 4640 | 1355 | 22537 |
| | Replace variable | 18224 | 3674 | 3996 | 25894 |
| $D_{meta-test}$ | Extract subclass | 4929 | 1302 | 205 | 6436 |
| | Move and rename class | 545 | 87 | 22 | 654 |
| | Rename class | 3287 | 557 | 147 | 3991 |
| | Extract and move method | 7273 | 1816 | 634 | 9723 |
| | Extract variable | 4744 | 1587 | 378 | 6709 |

### 5.3. Results and Description

In this paper, we have utilized accuracy, precision, and recall as evaluation metrics to assess our model and compare the results of conducted experiments. The reported results are the average metric values across one to 30 few-shot tasks for the meta-test stage. We analogize the results of our meta-learning model to those announced by Aniche et al. They have also reported their results based on evaluation metrics same as us. Aniche et al. have trained six machine learning and AI-based models (i.e., Neural Network, Random Forest, Decision Tree, Naïve Bayes, Support Vector Machine, and Logistic Regression) for classification of each of the 20 refactorings aforesaid in Table 1; in other words, six models for each refactoring and totally 120 models. On the other hand, we have solely trained *one model* for the classification of all referred refactorings in Table 1.

In the meta-learning context, it is popular that the evaluation metrics (accuracy, etc.) are reported as merely one averaged value for the whole of a dataset [28, 31, 32]. It is wildly different from traditional classification methods (Random Forest, etc.) that evaluation metrics are announced for each class of dataset individually. Accordingly, the results must be aligned and reconciled to compare our meta-learning consequences with those of Aniche et al.. To this end, we have calculated the weighted average of the reported accuracies by Aniche et al. for each model (Random Forest, etc.). The weights are equivalent to the ratio of samples of a particular refactoring to the total samples in *the $D_{meta-test}$*.

Table 3 demonstrates the results of seven machine learning methods on the same dataset. As is conspicuous, meta-learning has gained the best consequences compared to all traditional classification methods. The best accuracy in traditional methods belongs to Random Forest and the worst is related to Naïve Bayes, which are respectively about 2% and 23% lower than meta-learning accuracy. The precision and recall of meta-learning are also superior. The best precision in traditional methods belongs to Random Forest, which is 1% lower than our model. The Naïve Bayes model also has attained the same recall as our meta-learning model.

**Table 3.** Average test accuracy, precision, and recall of various machine learning models on $D_{meta-test}$. The results of our meta-learning model are represented in the last column. Numbers in bold characterize the top results.

|  | Logistic Regression | SVM (linear) | Naive Bayes (gaussian) | Decision Tree | Random Forest | Neural Network | Meta-Learning (ours) |
|---|---|---|---|---|---|---|---|
| **Precision** | 0.77 | 0.77 | 0.60 | 0.84 | 0.89 | 0.81 | **0.92** |
| **Recall** | 0.86 | 0.88 | 0.92 | 0.81 | 0.84 | 0.88 | **0.92** |
| **Accuracy** | 0.81 | 0.80 | 0.66 | 0.83 | 0.87 | 0.83 | **0.91** |

**Table 4.** Average test accuracy, precision, and recall on 2-way, 3-way, and 5-way classification

|  | 2-way | | | 3-way | | | 5-way | | |
|---|---|---|---|---|---|---|---|---|---|
|  | Precision | Recall | Accuracy | Precision | Recall | Accuracy | Precision | Recall | Accuracy |
| **3-shot** | 0.79 | 0.73 | 0.82 | 0.89 | 0.89 | 0.74 | 0.87 | 0.96 | 0.56 |
| **5-shot** | **0.92** | 0.92 | **0.91** | **0.92** | 0.90 | 0.81 | 0.90 | **0.97** | 0.72 |
| **10-shot** | 0.68 | 0.67 | 0.82 | 0.87 | 0.95 | 0.78 | 0.91 | 0.90 | 0.71 |

The reported results of meta-learning in Table 3 pertain to 2-way 5-shot classification. We have chosen this few-shot setting because it has revealed the best results among other settings. The accuracy, precision, and recall of nine different few-shot settings are detailed in Table 4. Obviously, 2-way 5-shot classification has achieved the best accuracy and precision, respectively equal to 0.91 and 0.92. The precision of the 3-way 5-shot classification is also 0.92. Furthermore, 5-way 5-shot classification has represented the best recall among others. Generally, the 5-shot setting for all 2-way, 3-way, and 5-way classifications has revealed the best result in our meta-learning model.

It is noteworthy that in traditional classification methods, the amount of training data is much more than adaptation data in meta-learning. As demonstrated in Table 2, the refactoring classes in $D_{meta-test}$ comprise 654 to 9723 data instances. Aniche et al. have trained their traditional classification models on all of these instances. While as is apparent in Figure 3, our meta-learning model has achieved its best results on merely one batch of tasks (each batch contains 25 tasks). Figure 3 illustrates the accuracy, precision, and recall vs. the number of batches for adaptation (meta-test stage) for 2-way and 3, 5, and 10-shot classification. As is evident, after the first batch for 5 and 10-shot and after two batches for 3-shot, all three metrics have revealed the best values roughly steady over 30 batches.

The cross-entropy loss function depicted in Figure 4 also has reached the lowest values after adaptation with just one or two batches of tasks. This Figure demonstrates the loss function values for 30 batches for 2-way and 3, 5, and 10-shot classification. The lowest loss is associated with 2-way 5-shot classification that is floating between 0.21 and 0.31 for different batches (as the highest accuracy, precision, and recall belong to 2-way 5-shot) and the highest loss is pertained to 2-way 10-shot, i.e., between 0.31 and 0.41 (as 2-way 10-shot possesses the lowest accuracy, precision, and recall).

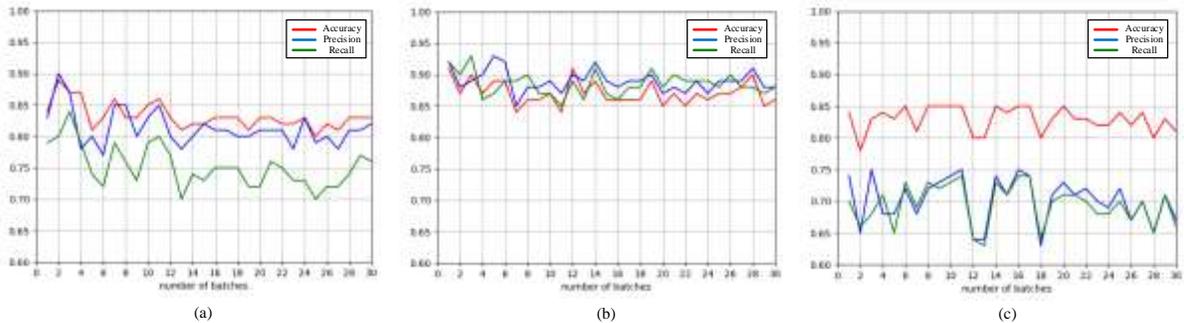

**Figure 3.** Accuracy, precision, and recall vs. the number of batches for 2-way and (a): 3-shot, (b): 5-shot, (c): 10-shot meta-learning classification.

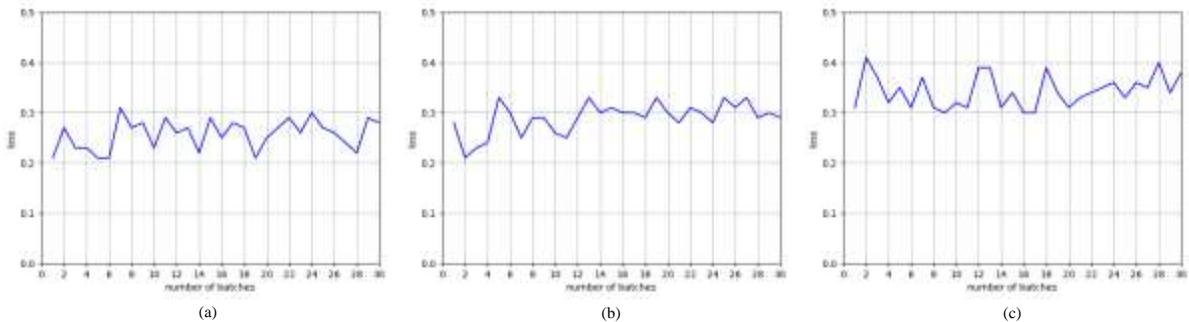

**Figure 4.** The value of cross-entropy loss function for 2-way and (a): 3-shot, (b): 5-shot, (c): 10-shot meta-learning classification.

The results of conducted experiments in this section shed light on the effectiveness and competitiveness of the meta-learning approach in software refactoring. Our meta-learning model has gained the best accuracy, precision, and recall among the seven classification methods. Due to the fine-tuning of the meta-trained model to a model for the classification of unseen types of refactoring (refactorings in $D_{meta-test}$) with solely one or two batches of tasks, it is pragmatic to utilize the meta-trained model for the classification of more upcoming refactoring types (other than refactoring types introduced in Table 1) with few amounts of data too. As illustrated in this section, the meta-learning approach can train a versatile model with fruitful results because of spending few annotated data, processing time, and power compared with traditional machine learning methods.

## 6. Conclusion and Future Works

In this study, we proposed a meta-learning approach for software refactoring. We trained an MLP network with three hidden layers on a dataset containing an abundance of refactoring data and then adapted the trained model for the classification of refactorings with the adversity of data scarcity. The wealth of our approach is due to the consumption of a few annotated data, processing time and power for adaptation, as well as the feasibility of adaptation for classification of any upcoming refactoring type. Our model revealed the highest accuracy, precision, and recall, respectively equal to 0.91, 0.92, and 0.92 from six traditional classification methods: Neural Network, Random Forest, Decision Tree, Naïve Bayes, Support Vector Machine, and Logistic Regression on the same dataset.

As previously stated, training and fine-tuning (testing) data must be drawn from related distributions in meta-learning and generally knowledge-transfer approaches. Accordingly, in the future, it is plausible to apply meta-learning on any software-related task, such as anti-pattern, bad smell, and design pattern detection in software quality assurance and maintenance, because it is realistic to assume that all of these data come from related and analogous distributions. However, it is worthy of mention that one significant assumption is accessibility to a dataset with comprehensive annotated data for training the meta-learner. Due to adequate annotated data for some software-related tasks and the lack of such data for others, meta-learning can offer an applicable technique to deal with data-driven problems in the software area.